# The Laser Ablation Production of Platinum Nitride and its Possible Structure.

G. Soto and M.G. Moreno-Armenta

*Universidad Nacional Autónoma de México*

*Km 107 carretera Tijuana-Ensenada, Ensenada Baja California, México.*

**Abstract**

The synthesis of platinum nitride by the laser ablation method is reported. The spectroscopic results show that nitrogen is in interstitial sites of platinum as N-units, contradicting the accepted configuration for $PtN_2$ where it is as $N_2$-units. To elucidate this point we did density functional calculations to correlate composition with nitrogen sites. For dilute nitrogen concentrations, $x < 0.2$, nitrogen would be in six-fold coordinated sites (octahedral interstices) as N-units. For $0.2 < x < 1.5$, nitrogen would be in tetrahedral interstices as N-units. Only for $x > 1.5$ the $N_2$-configuration in octahedral sites is attained.

**Corresponding Author:** **G. Soto.**

**Universidad Nacional Autónoma de México.**

**P.O. Box 439036, San Ysidro, CA**

**92143-9036, USA**

**Tel: +52+646+1744602, Fax: +52+646+1744603**

**E-Mail:** gerardo@ccmc.unam.mx

In past times there was a widespread belief that the noble metals do not develop nitrides, subsequently the recent synthesis of platinum nitride is of significance for the scientific community. Gregoryanz *et al* shows that platinum nitride can be prepared by high-pressure high-temperature conditions [1]. Simultaneously we show the deposition of platinum nitride films by the Reactive Pulsed Laser Deposition (RPLD) method [2]. This last synthesis was recently revised by us. Figure 1 shows the corresponding Auger spectrum. This is an undeniably confirmation that nitrogen can be chemisorbed within the platinum lattice to produce the alloy. The [N] / [Pt] ratio is lower than that of the high pressure methods [1, 3-4]. It is then concluded that there are many compositions for platinum nitride, and it has to be an interstitial alloy [5]. The interstitial alloys result when large atoms are combined with much smaller atoms that are not too electropositive. The small atoms occupy the interstices in the *close-packed* metal structures. In our opinion the most remarkable aspects in the synthesis by Gregoryanz is the atypical arrangement taken by nitrogen within the metallic lattice [1, 3]. In interstitial alloys the nonmetal atom is positioned in interstitial sites of the parent metal as single units; instead in $PtN_2$ nitrogen adopt an unusual dimer configuration ($N_2$-units). Similar anions have been found in $SrN_2$ and in $BaN_2$ [6], inclusive $N_3$-chains are present in $AgN_3$ [7], but to our understanding there are not reports of similar happening in closed-packed metallic structures. In the laser-ablation synthesis the x-ray photoelectron spectroscopy (XPS) show that nitrogen is as $N_1$-units (insert of figure 1). $N_1$ can easily be set apart of $N_2$ since the $N_2$ species have a noticeable splitting of the 2s energy-level due to the formation of $\sigma_g$ and $\sigma_u$ anion-molecular states, as it is seen in figure 4 of reference 4 and figure 2 of reference 8. Since this splitting does not appear in the N-*2s* transition we conclude that nitrogen is as $N_1$-units. This is in conflicts with the accepted configuration for platinum nitride.

The properties of interstitial alloys are dependent on the preparation conditions due to the different structures and compositions that they can take up [5]. With the aim to elucidate this point we did a series of density functional calculations for platinum nitride(s). The method consists in creating a supercell of Pt-atoms in *cubic close-packet* configuration and filling its interstitial holes with N-atoms in sequential manner. The N to Pt ratio is related to the number of interstices used. It is covered the occupancy of octahedral interstices, tetrahedral interstices and '*pyrite-like'* interstices. The pyrite-like sites are defined as octahedral holes occupied by $N_2$ entities. The trends in energy as a function of composition of such hypothetical structures will be a sign of the most probable nitrogen occupation sites. The probable sites are related to physically realizable structures.

The supercell created for calculations was with eight Pt-atoms, which are associated with eight octahedral interstices and sixteen tetrahedral interstices. The compositional range (*x*) goes to a maximum $PtN_{x=1}$ for octahedral occupancy when all these sites are used by N-monomers, and to $PtN_{x=2}$ when used by $N_2$-dimers. For the tetrahedral occupancy with N-monomers the composition reach a maximum of $PtN_{x=2}$. Platinum nitride is assumed to be of cubic symmetry; despite that in this work the symmetry was reduced down to rhombohedral (hexagonal) symmetry to allow for relaxation in volume and *c/a*. The relaxation in hexagonal coordinates in *c*-direction is equivalent to relaxation in the <111> direction in cubic coordinates. In this way we avoid the mechanical instabilities associated to some structures [4, 9-10]. The numerical calculations were carried out with the Wien2k code [11]. Calculations were made using the Generalized Gradient Approximation (GGA) of Perdew *et al* [12, 13]. The muffin tin radii for all calculations were done with 2.0 Bohr for Pt and 1.2 Bohr for N, respectively. The calculations have been done using an $R_{mt}K_{max}$ ~ 7 and $G_{max}$ ~ 12. We take an energy cutoff of -8.0 Ry to separate the core from the valence states. The number of *k*-points in the irreducible wedge of the Brillouin zone was 44. The energy convergence was set to 0.00001 Ry. The equilibrium volume, cohesive energy and bulk modulus was obtained after fitting the equation of state of Murnaghan to the *E* vs. *Vol* curves [14].

The close-packed structures contain two types of holes. The tetrahedral hole is surrounded by four metal atoms at the corners of a tetrahedron. An atom will fit into this hole as long as its radius is no more than $0.225r_m$, where $r_m$ is the radius of the metal atom. The second type of hole is surrounded by six atoms, determining an octahedron. An atom will fit into this hole as long as its radius is no more than $0.414r_m$. The metallic radius of Pt is 138.7 pm, while the covalent radius of nitrogen is 70.2 pm [15]. Evidently, N does not fit in any of the interstitial holes of Pt. A cell expansion is a required for nitride formation, as it is show in Figure 2, where is plotted the change in cell volume as a function of nitrogen assimilation when $N_1$ is into octahedral sites; $N_1$ is tetrahedral sites; and $N_2$ in octahedral (pyrite-like) sites. In this plot we have included the calculated cell volumes for platinum nitride in zincblende, rocksalt, fluorite and pyrite structures as they have been reported (dissimilar symbols in plot). The occupation of tetrahedral interstices would require a larger cell expansion than the occupation of octahedral interstices for the same composition. This is consequence of the smaller space available for tetrahedral configuration. Very interesting, the use of octahedral sites by $N_2$-entities will require smaller cell expansion than the use of tetrahedral sites by $N_1$-entities. The pyrite-like platinum nitrides have a more efficient atom-packaging than the fluorite-like platinum nitrides.

In figure 3 we show side-by-side the bulk modulus and the valence band electron concentration as a function of *x* in $PtN_x$. The intention of this plot is to show the strong correlation that exists between bulk modulus and electronic concentration. The occupation of the Pt holes produces a very big decrease of electronic concentrations due to the acute cell expansion required for nitride formation (figure 2), being most perceptible for low nitrogen contents. However, as the N to Pt ratio is incremented the number of electrons provided by N overpasses the negative effect of cell expansion. As consequence the electronic concentration is recovered, exceeding by far the original electron concentration of Pt in the pyrite-like phases. Following the trends of electron concentration is the bulk modulus. The use of pyrite-like sites by $N_2$ produces larger bulk moduli than the use of tetrahedral or octahedral sites by $N_1$, this when $x \rightarrow 2$. However for low nitrogen concentration (*x* near to 1), the larger bulk modulus is with $N_1$ in octahedral sites.

The important question to solve in this work is: what are the most probable sites to be occupied by nitrogen in platinum nitride for any composition? The answer can be given by the alloy energies, defined as [16]:

$$E_i^{alloy} = \frac{E_i^{Tot} - E_{Pt}^{Tot} - nE_{Nf}}{m+n} \tag{4}$$

Where $E_i^{Tot}$ is the total calculated energy of the *i*-cell, which contains *m* and *n* number of Pt and N atoms. $E_{Pt}^{Tot}$ is the total energy of *m* atoms of Pt in its c*cp* ground state (as calculated). $E_{Nf}$ is the energy of nitrogen in free atomic state. To compare energies at equal volumes is convenient to employ the energy density, as it is plotted in figure 4 along with the alloy energy. It seems that for $x < 1.5$ nitrogen would be assimilated preferably as $N_1$-units since they have lower (more stable) alloying energies. This is in agreement with our previous conclusion by means of XPS. However for $x > 1.5$ there a clear margin favorable to the $N_2$ dimer configuration in octahedral interstices, explaining the syntheses and interpretations made by Gregoryanz and Crowhurst [1, 3]. Additionally, for very dilute nitrogen concentrations, $x < 0.2$, nitrogen would be preferably in six-fold coordinated sites as $N_1$-units (octahedral interstices). For the $0.2 < x < 1$ interval the result of the current calculations are not decisive. For example, at the composition $x = 1$ the energy is favorable by a tiny margin to octahedral configuration (NaCl-like). Conversely, at $x = 0.5$ the calculations indicate that the use of tetrahedral holes is favored in any condition.

In summary, even if the general description of interstitial alloys suggests that the nonmetal atom simply occupy the interstices in the metal lattice, this is an oversimplified visualization. The calculation performed here indicates that there are many possible structures for platinum nitrides, and the adopted structures will change as a function of composition. We see here that the most viable structures when the nitrogen composition excess a limit of 1.5 is with the N-atoms in dimer configuration. For $x$ = 2 this gives the reported pyrite-like $PtN_2$ phase. However for low nitrogen concentrations N will be as $N_1$-units. The use of octahedral or tetrahedral sites depends on N-concentration. The changes in composition and structures in $PtN_x$ will result in different electronic band structures (not show here), which goes from metallic to dielectric. In this way the divergent properties found for platinum nitride produce by two different methods can be explained.

There are many additional structural possibilities worth to be explored that are not covered here. For example, the rearrangement of the metallic lattice in different stacking sequences is very common in heavy metal nitrides, or the creation of metallic vacancies, which is the case in copper nitride [17].

The authors are grateful to C. González, D. Garcia and J. Peralta for technical assistance. This work was supported by Supercomputer Center DGSCA-UNAM and DGAPA project IN120306.

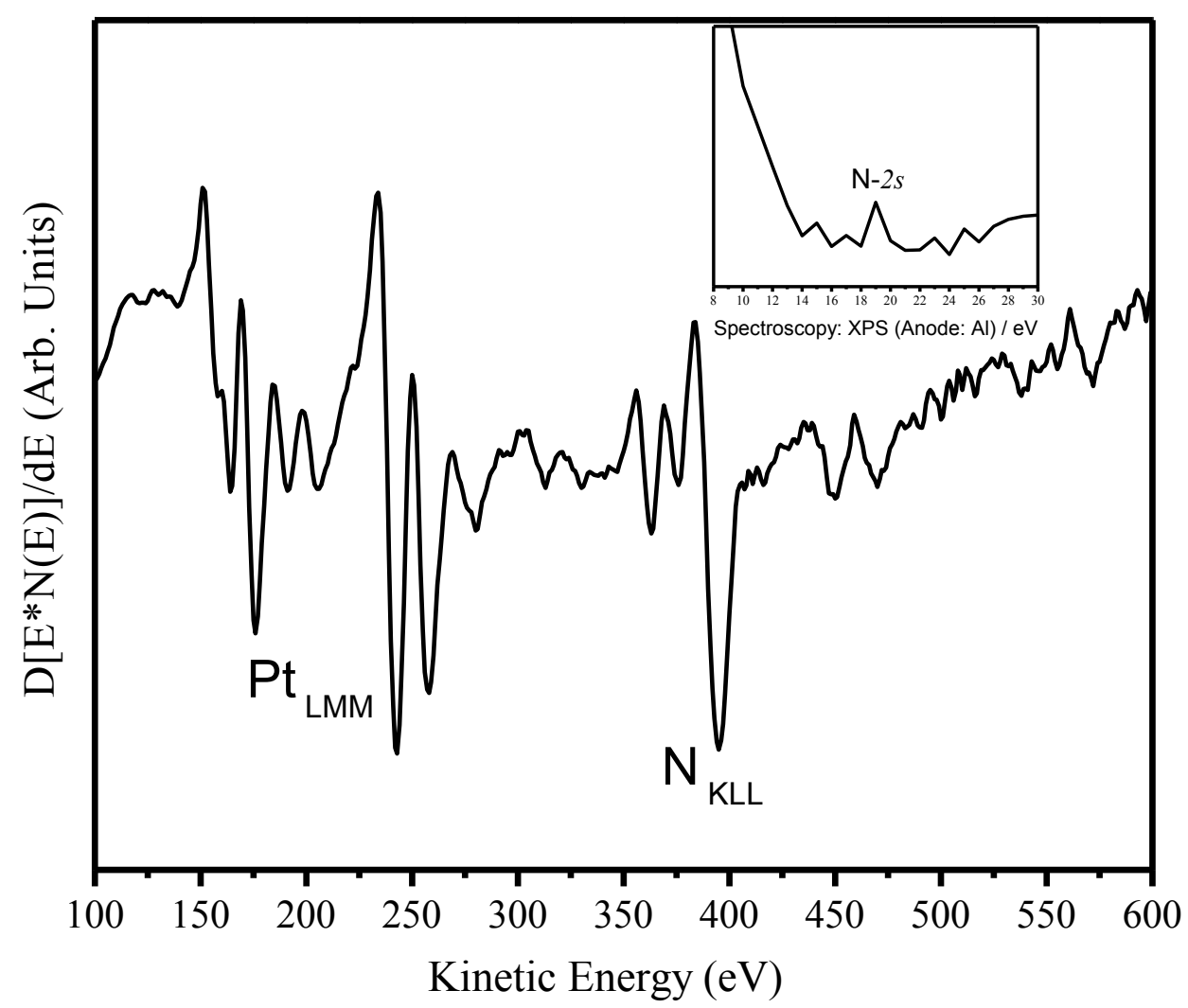

Figure 1.- Auger spectra of platinum nitride films. The insert plot shows a typical x-ray photoelectron spectrum in the valence band for $PtN_x$.

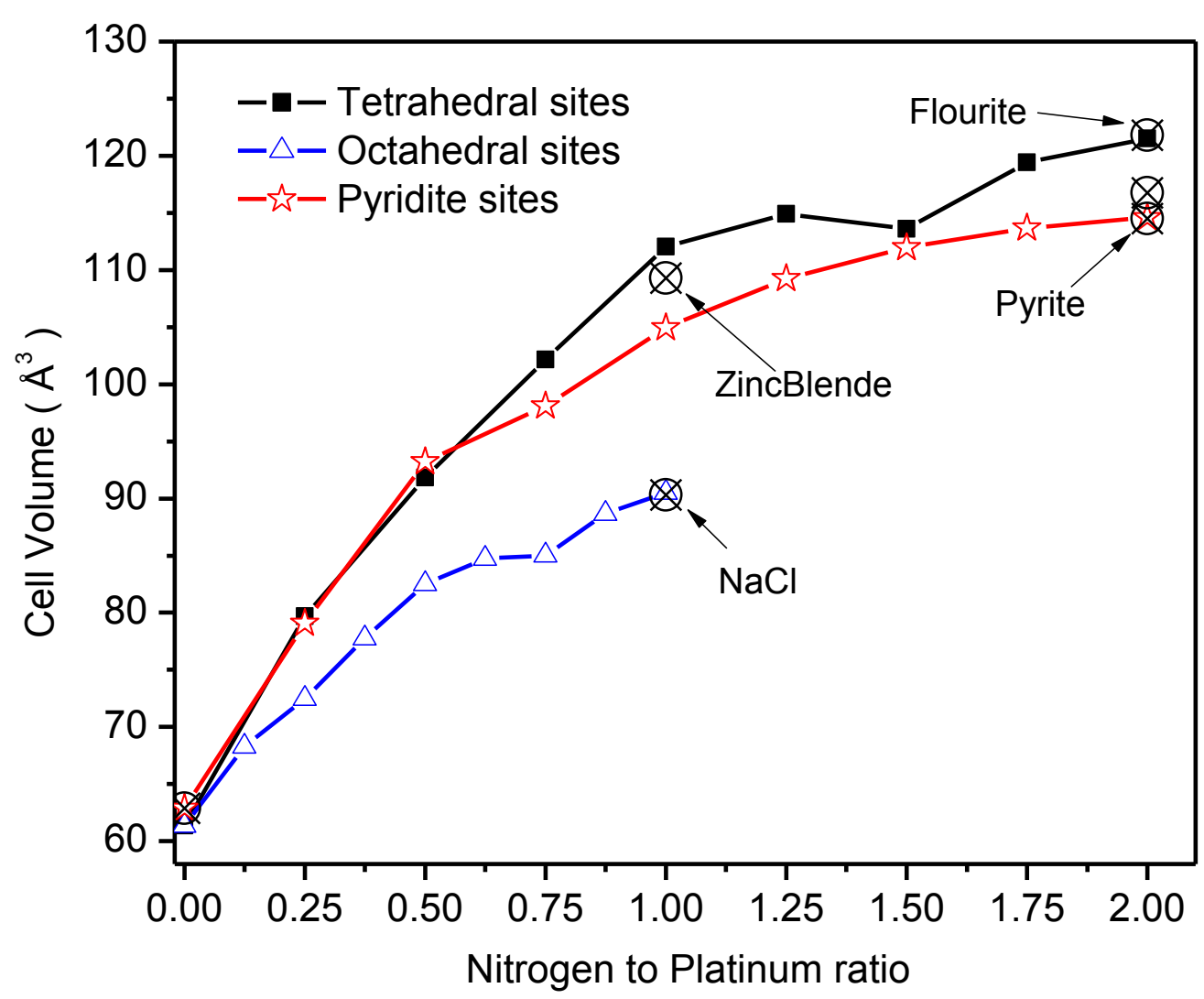

Figure 2.- Cell volume as a function of composition for $PtN_x$ with nitrogen in octahedral sites (Triangles), tetrahedral sites (squares) and pyrite-like sites (stars). Also it is shows the cell volume in PtN composition (NaCl and Zincblende) and $PtN_2$ composition with Flourite and Pyrite structures.

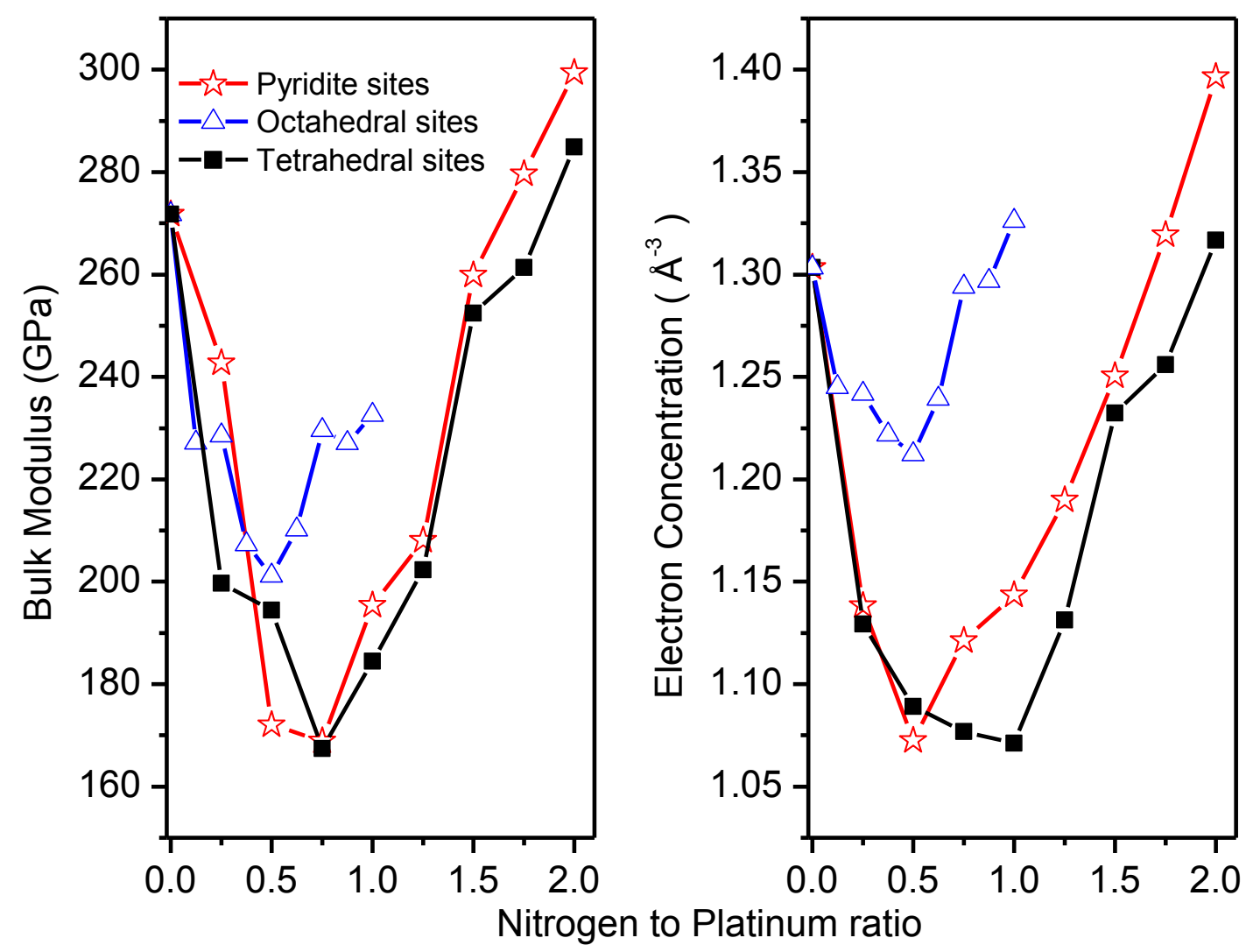

Figure 3.- Bulk modulus and electronic concentration as a function of composition

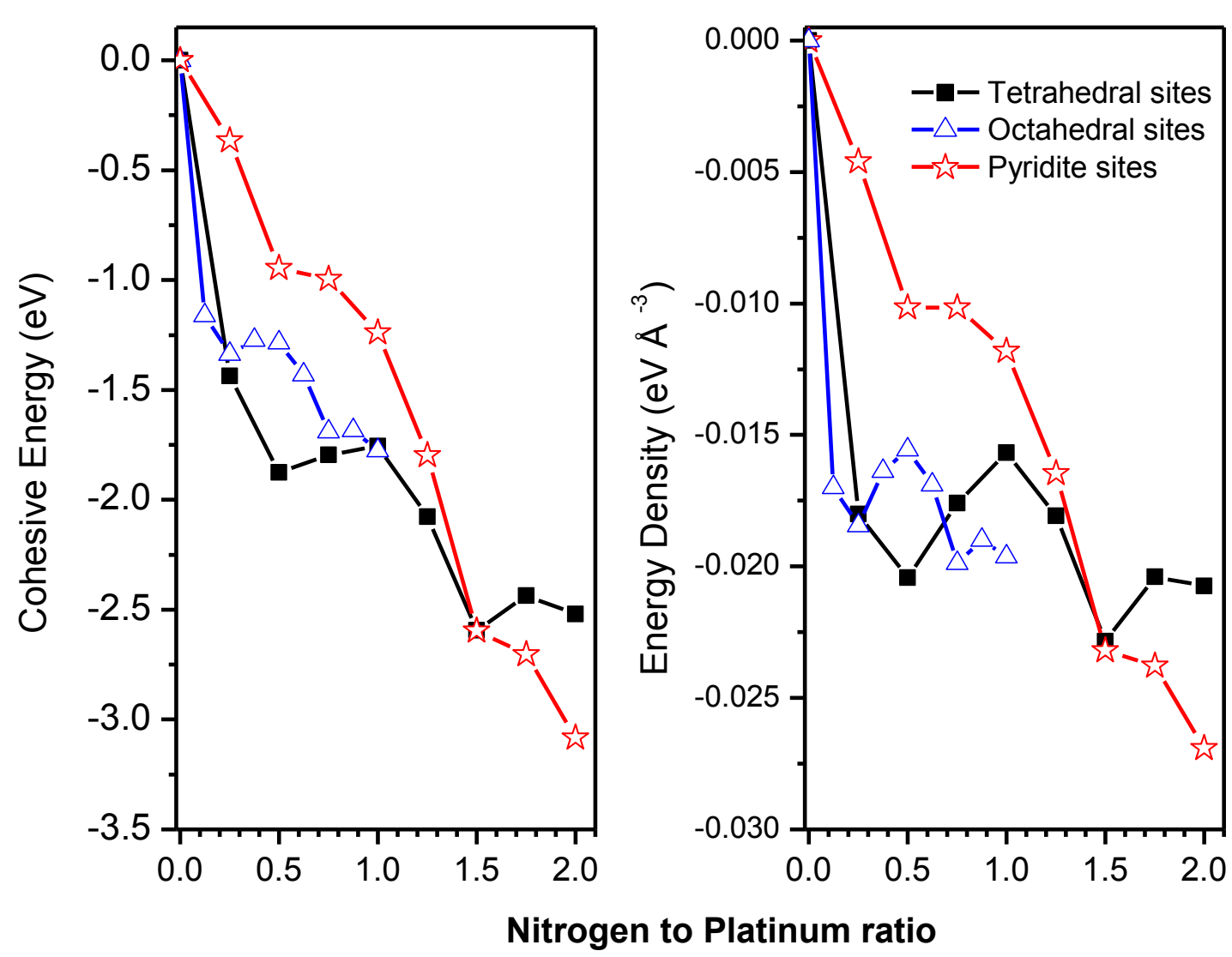

Figure 4.- Cohesive energy and energy densities as a function of composition for $PtN_x$.